\documentclass[aps,prl,twocolumn,superscriptaddress]{revtex4}

\usepackage{bm}
\usepackage{amsmath}
\usepackage{amssymb}
\usepackage{graphicx}
\usepackage{color}

\input{epsf}
\begin{document}
\title{Shuttling of Spin Polarized Electrons in Molecular Transistors}

\author{O. A. Ilinskaya}
\affiliation{B.~Verkin Institute for Low Temperature Physics and
Engineering of the National Academy of Sciences of Ukraine, 47 Lenin
Ave., Kharkiv 61103, Ukraine}

\author{S. I. Kulinich}
\affiliation{B.~Verkin Institute for Low Temperature Physics and
Engineering of the National Academy of Sciences of Ukraine, 47 Lenin
Ave., Kharkiv 61103, Ukraine} \affiliation{Department of Physics,
University of Gothenburg, SE-412 96 G{\" o}teborg, Sweden}
\affiliation{Department of Physics and Astronomy, Seoul National
University, Seoul 151-747, Korea}

\author{I. V. Krive}
\affiliation{B.~Verkin Institute for Low Temperature Physics and
Engineering of the National Academy of Sciences of Ukraine, 47 Lenin
Ave., Kharkiv 61103, Ukraine} \affiliation{Department of Physics,
University of Gothenburg, SE-412 96 G{\" o}teborg, Sweden}
\affiliation{Department of Physics and Astronomy, Seoul National
University, Seoul 151-747, Korea} \affiliation{Physical Department,
V.~N.~Karazin National University, Kharkiv 61077, Ukraine}

\author{R. I. Shekhter}
\affiliation{Department of Physics, University of Gothenburg, SE-412
96 G{\" o}teborg, Sweden}

\author{Y. W. Park}
\affiliation{Department of Physics and Astronomy, Seoul National
University, Seoul 151-747, Korea}

\author{M. Jonson\footnote{Corresponding author. E-mail: mats.jonson@physics.gu.se}}
\affiliation{Department of Physics, University of Gothenburg, SE-412
96 G{\" o}teborg, Sweden} \affiliation{SUPA, Institute of Photonics
and Quantum Sciences, Heriot-Watt University, Edinburgh, EH14 4AS,
Scotland, UK}
\date{\today}
\begin{abstract}
Shuttling of electrons in single-molecule transistors with
magnetic leads in the presence of an external magnetic field is
considered theoretically. For a current of partially spin-polarized electrons a shuttle 
instability 
is predicted to occur for 
a finite interval of external magnetic field
strengths. The lower critical magnetic field is determined by the
degree 
of spin polarization and it vanishes 
as the spin polarization approaches 100\%. The feasibility of
detecting magnetic shuttling in a $C_{60}$-based molecular
transistor with magnetic (Ni) electrodes is discussed
[A.~N.~Pasupathy et al., Science {\bf 306}, 86 (2004)].\\

\noindent
Keywords: single-electron shuttling, molecular transistors, spintronics
\end{abstract}

\maketitle

\section{1. Introduction}
In recent years the effect of the spin of electrons on the
transport properties of nanostructures have been studied
intensively, both theoretically and experimentally. In the context
of spin-based electronics (spintronics) the possibility to control
electrical currents by a weak external magnetic field using the
Zeeman and/or the spin-orbit interaction is one of the main goals.

Magnetic materials and especially half-metals are natural sources of
spin-polarized electrons for spintronics. Transport of spin-polarized
electrons in nanostructures (quantum dots, suspended nanowires,
etc.) in external magnetic field results in new phenomena where spin,
charge and mechanical degrees of freedom are strongly inter-related.
In this new field of investigations (spintromechanics, see
Ref.~\onlinecite{pulkin}) the presence of a mechanically ``soft"
subsystem results both in a strong enhancement of spintronic effects
and in magnetic control of the mechanical subsystem in the
classical as well as in the quantum transport regimes.

Vibrational effects are known to be important for the transport
properties of molecular transistors (see, e.g., the reviews in
Refs.~\onlinecite{galperin} and \onlinecite{krive}). In single-molecule
transistors a strong electron-vibron coupling was observed
in a $C_{60}$-based transistor with nonmagnetic (gold)
leads \cite{park}. The measured current-voltage characteristics in
this experiment revealed low-energy periodic step-like features.
They were interpreted as a signature of vibron-assisted electron
tunneling via the fullerene molecule. Experimental $I-V$ curves were
theoretically explained \cite{shekhter,flensberg} in the frames of a
simple model of a single-level quantum dot strongly coupled to a
single vibrational mode and weakly coupled to the source and drain
electrodes.

Later on $C_{60}$-based molecular transistors with magnetic (Ni)
leads were fabricated \cite{pasupathy}. In samples where the
tunneling coupling to the ferromagnetic electrodes were
relatively strong ($\sim$ tens of meV), Kondo-assisted tunneling via the
$C_{60}$ molecule was observed. These measurements also proved the
presence of a strong inhomogeneous magnetic field produced by the
ferromagnetic electrodes in the nano-gap between them. In samples with weak
tunneling couplings the usual Coulomb blockade picture for a single-electron 
transistor was observed.

In the present paper we formulate the conditions for the
appearance of a vibrational instability of a fullerene molecule
suspended in the gap between two magnetic leads with opposite
magnetization. Electron shuttling of spin-polarized electrons
produced by magnetic (exchange) forces was
predicted in Ref.~\onlinecite{kul} for the case of 100\%
polarization of the leads. In this limit (realized for
half-metals) the electric current is blocked (spin blockade) in
the absence of spin-flips induced by, e.g., an external magnetic
field. It was shown that in the absence of dissipation in the
mechanical subsystem such a magnetic field triggers a shuttle
instability even for vanishingly small fields \cite{kul}. In the
presence of dissipation a threshold magnetic field is determined
by the rate of dissipation and it is small for a weak dissipation.

One of
our aims here is to develop a theory of
magnetic shuttling for conditions corresponding to the experimental
set-up of Ref.~\onlinecite{pasupathy}, where the electrons in the
ferromagnetic leads were partially polarized ($\sim 30\%$). The
absence of a spin blockade in this case qualitatively changes the criterion for
electron shuttling. We will show that even in the absence of mechanical
dissipation, a shuttling regime of electron transport occurs in a
finite interval of external magnetic field strengths,
$H_{\text{min}}<H<H_{\text{max}}$, where $H_{\text{min}}$ is
determined by the degree of spin polarization $\eta$.

In particular, for a high degree of spin polarization
$(\eta\rightarrow 1)$ and if $\Gamma\gg\hbar\omega$ (where $\Gamma/\hbar$
is the tunneling
rate of majority spin electrons and $\omega/2\pi$ is the
mechanical vibration frequency of the fullerene) the threshold magnetic field for
reaching the shuttling regime of electron transport reads:
$H_{\text{min}}\sim\sqrt{1-\eta}\,\Gamma$. In the limit of ``hard"
vibrons, $\Gamma\ll\hbar\omega$, the threshold field is determined
by the vibron energy, $H_{\text{min}}\sim\sqrt{1-\eta}\,\hbar\omega$.

The calculations outlined below aim at determining the rate of change
$r(H)$ (to be defined later) of the
center-of-mass coordinate of the single-molecule shuttle, the sign of
which then allows us to formulate the
conditions required to observe shuttling of spin-polarized
electrons in $C_{60}$-based molecular transistors.

\section{2. Hamiltonian and Equations of Motion}
The Hamiltonian of a magnetically driven single electron shuttle (see
Refs.~\onlinecite{kul} -- \onlinecite{parafilo}) consists of four terms,
\begin{equation}\label{4}
\hat H=\sum_{j=S,D}\left(\hat H_j+\hat H_{t,j}\right)+\hat H_d+\hat
H_v,
\end{equation}
where $\hat H_j$ is the standard Hamiltonian of noninteracting
electrons in the source ($j=S$) and drain ($j=D$) electrodes, $\hat
H_{t,j}$ is a tunneling Hamiltonian with coordinate dependent
tunneling amplitudes $t_j(\hat x)$ and $\hat H_d$ is the Hamiltonian
of a single level ($\varepsilon_0$) quantum dot (QD) magnetically coupled to
leads of spin-polarized electrons by coordinate dependent exchange
interactions $J_j(\hat x)$. It has been shown \cite{krive} that the shuttling
regime of single-electron transport can be realized in the presence
of an external magnetic field $H_{\text{ext}}$ for oppositely
magnetized source and drain electrodes. If 
the external
magnetic field is directed perpendicular to the antiparallel polarization vectors
of the leads, the QD Hamiltonian reads
\begin{eqnarray}\label{5}
&&\hat H_d= \left[\varepsilon_0-\frac{J(\hat x)}{2} \right]
a^\dagger_\uparrow a_\uparrow+\left[\varepsilon_0+\frac{J(\hat
x)}{2} \right] a^\dagger_\downarrow
a_\downarrow\nonumber\\&&\hspace{0.65cm}-\frac{g\mu
H}{2}\left(a^\dagger_\uparrow a_\downarrow+a^\dagger_\downarrow
a_\uparrow\right)+Ua^\dagger_\uparrow a_\uparrow
a^\dagger_\downarrow a_\downarrow.
\end{eqnarray}
Here $a^\dag_\sigma$ $(a_\sigma)$ is the creation (annihilation)
operator for an electron with spin projection $\sigma=\uparrow,\,
\downarrow$ on the dot, $J(\hat x)=J_S(\hat x)-J_D(\hat x)$, $\mu$
is the Bohr magneton ($g$ is the gyromagnetic ratio), and $U$ is the
Coulomb repulsion energy. QD vibrations are described by the
harmonic oscillator Hamiltonian
\begin{equation}\label{6}
\hat H_v=\frac{\hat p^2}{2m}+\frac{m\omega^2\hat x^2}{2},
\end{equation}
where $\hat x$ is the displacement operator, $\hat p$ is the
canonical conjugated momentum ($\left[\hat x, \hat p
\right]={i}\hbar$), $m$ is the mass and $\omega$ is the
(angular) vibration frequency of the QD.

The aim of the present paper is to find the conditions under
which magnetically driven shuttling in experiments with
fullerene-based single-molecule transistors \cite{park,pasupathy}
can be realized.
In this case the ferromagnetic leads are characterized by a certain
degree $0<\eta<1$ of spin polarization ($\sim 30\%$ in the
experiment of Ref.~\onlinecite{pasupathy}) and the characteristic
vibron energy $\hbar\omega$ ($\sim$ several meV,
Ref.~\onlinecite{park}) is larger or of the order of the energy
scale $\Gamma$ that characterizes the tunneling coupling to the leads,
$\Gamma\leq\hbar\omega$.

It follows from the above considerations that we are not in the adiabatic regime
of mechanical motion  (which is said to be antiadiabatic when
$\Gamma\ll \hbar\omega$) and the physical picture of ``magnetic
shuttling" developed in Ref.~\onlinecite{kul} (see also
Ref.~\onlinecite{parafilo}) for the adiabatic regime does not hold
here. As in Ref.~\onlinecite{ilin} we will solve the problem using
equations of motion for the reduced density operator. In the limit
$eV\gg\Gamma,\, \hbar\omega,\, \mu H,\, k_BT$ ($T$ is the
temperature, $V$ is the bias voltage) the density operator can be
factorized into the product of a QD density operator and an
equilibrium density matrix for the leads. In the Coulomb blockade
regime, $eV,\,T\ll U$, the matrix elements of the QD density
operator ($\rho_0=\langle 0\vert\hat \rho_d\vert0 \rangle$,
$\rho_\sigma=\langle \sigma\vert \hat \rho_d\vert\sigma\rangle$,
$\rho_{\uparrow\downarrow}=
\langle\uparrow\vert\hat\rho_d\vert\downarrow\rangle$) in the
Hilbert space of a singly occupied dot level (they are still
operators in the Hilbert space of a harmonic oscillator) are
determined by the following set of equations \cite{mis},
\begin{eqnarray}\label{baseq}
&&\frac{\partial \rho_0}{\partial t}=-i\left[ H_v,
\rho_0\right]-\frac{1}{2}\left\{\Gamma_S^\uparrow(\hat
x)+\Gamma_S^\downarrow(\hat x),
\rho_0\right\}\nonumber\\
&&\hspace{0.7cm}\label{baseq1} +\sqrt{\Gamma_D^\uparrow(\hat
x)}\rho_\uparrow\sqrt{\Gamma_D^\uparrow(\hat x)}
+\sqrt{\Gamma_D^\downarrow(\hat x)}
\rho_\downarrow\sqrt{\Gamma_D^\downarrow(\hat x)},\\
&&\frac{\partial \rho_\uparrow}{\partial t}=-i\left[
H_v,\rho_\uparrow\right]+\frac{{i}}{2}\left[J(\hat x),
\rho_\uparrow\right]+ \frac{{i} h}{2}
\left(\rho_{\uparrow\downarrow}
-\rho^{\dag}_{\uparrow\downarrow}\right)
\nonumber\\&&\hspace{0.7cm}+\sqrt{\Gamma_S^\uparrow(\hat x)}\rho_0
\sqrt{\Gamma_S^\uparrow(\hat x)}-
\frac{1}{2}\left\{\Gamma_D^\uparrow
(\hat x), \rho_\uparrow\right\},\\
&& \frac{\partial \rho_\downarrow}{\partial t}=-i\left[
H_v,\rho_\downarrow\right]-
\frac{{i}}{2}\left[J(x),\rho_\downarrow\right]-\frac{{i}
h}{2}\left(\rho_{\uparrow\downarrow}-\rho^\dag_{\uparrow\downarrow}
\right)\nonumber\\
&&\hspace{0.7cm}+\sqrt{\Gamma_S^\downarrow(\hat
x)}\rho_0\sqrt{\Gamma_S^\downarrow(\hat x)} -\frac{1}{2}
\left\{\Gamma_D^\downarrow(\hat x),\rho_\downarrow\right\},\\
&& \frac{\partial \rho_{\uparrow\downarrow}}{\partial t}= -i\left[
H_v, \rho_{\uparrow\downarrow}\right]+\frac{{i}}{2}\left\{J(\hat
x),\rho_{\uparrow \downarrow}\right\}+\frac{{i} h}{2}
\left(\rho_\uparrow-\rho_\downarrow\right)\nonumber\\&&\hspace{0.7cm}
-\frac{\rho_{\uparrow\downarrow} }{2}\Gamma_D^\downarrow(\hat x)
-\Gamma_D^\uparrow(\hat x)\frac{\rho_{\uparrow\downarrow}}{2}.
\label{baseq2}
\end{eqnarray}
In order to solve Eqs.~(\ref{baseq1}) -- (\ref{baseq2})
it is convenient to introduce dimensionless variables
for time ($t\omega\rightarrow t$), dot displacement ($\hat
x/x_0\rightarrow \hat x$, where $x_0=\sqrt{\hbar/m\omega}$ is the
zero-point oscillation amplitude), momentum ($\hat p
x_0/\hbar\rightarrow \hat p$) and for various characteristic
energies ($\hbar\omega\rightarrow 1$, $g\mu
H/\hbar\omega\rightarrow h$, $\Gamma_j^\sigma(\hat
x)/\hbar\omega\rightarrow\Gamma_j^\sigma (\hat x)$, where
$\Gamma_j^\sigma(\hat x)=2\pi\nu\vert t_{j,\sigma}(\hat x)\vert^2$
is the level width, $\nu$ is the density of states).

\section{3. Influence of Spin Polarization on the Magnetic Shuttle
Instability} We are interested in the conditions under which a
magnetic shuttle instability occurs. Therefore it is sufficient to
expand tunneling amplitudes and exchange energies up to linear
terms in the coordinate operator $\hat x$,
\begin{equation}\label{7}
J(\hat x)=J_0-\alpha \hat x,\,\Gamma_{S/D}^\sigma(\hat
x)=\Gamma_{S/D}^\sigma\left(1\mp\frac{\hat x}{l}\right),
\end{equation}
where $l$ is the characteristic electron tunneling length (of the
order of 1~{\AA}).

Note that the zero-point fluctuation amplitude of a fullerene molecule 
in the vicinity of the Lennard-Jones potential minimum is
much smaller than the electron tunneling length, $l\gg 1$. Since the
spatial scale of the exchange interaction is also determined by the
tunneling length $l$, we can treat the mechanical coordinate $\hat x$ as
a classical variable. The equation of motion for the classical
coordinate $x_c$ takes the form
\begin{equation}\label{8}
\frac{d^2 x_c}{d
t^2}+x_c=-\frac{\alpha}{2}\text{Tr}\left(\rho_\uparrow-
\rho_\downarrow\right)\,.
\end{equation}
The r.h.s. of Eq. (\ref{8}) can readily be found from the set of
linear equations (\ref{baseq1}) -- (\ref{baseq2}) when the operator $\hat
x$ is replaced by $x_c$.

To simplify analytical calculations we consider a symmetric junction for which
$J_0=0$, $\Gamma_S^\uparrow=\Gamma_D^\downarrow=\Gamma$, and
$\Gamma_S^\downarrow=\Gamma_D^\uparrow=\gamma$. To first order in
the small dimensionless parameter
$\alpha/l\sim\left(J/\hbar\omega\right) \left(x_0/l\right)^2\ll 1$
Eq.~(\ref{8}) is reduced to a linear equation. For the time
dependent coordinate displacement $x_1=x_1(t)$ this equation reads
\begin{equation}\label{9}
\frac{d^2
x_1}{dt^2}+x_1=-\frac{\alpha}{2l}\int_{-\infty}^tdt'\langle e_0\vert
e^{\hat A(t-t')}\vert e\rangle x_1(t').
\end{equation}
Here the matrix $\hat A$ takes the form
\begin{equation}\label{90}
\hat A=\frac{1}{2}\left(
\begin{array}{ccc}
-\Gamma_1&-\Gamma_2& -2h \\
\Gamma_2 &-3\Gamma_1 & 0 \\
2h & 0& -\Gamma_1
\end{array}\right),
\end{equation}
$\Gamma_{1,2}=\Gamma\pm\gamma$, $\vert e_0\rangle=(1,0,0)^T$, and the
vector $\vert e\rangle=(e_1,e_2,e_3)^T$ has the components
\begin{eqnarray}
&&e_1= -\frac{2\Gamma_2}{\Delta}(\Gamma_1^2-\Gamma_2^2),\,
e_2=-\frac{2\Gamma_1}
{\Delta}\left(\Gamma_1^2-\Gamma_2^2+4h^2\right) \nonumber\\
&&e_3=-\frac{4h\Gamma_1\Gamma_2}{\Delta},\,
\Delta=3\Gamma_1^2+\Gamma_2^2+12h^2.
\end{eqnarray}

A shuttle instability occurs when the oscillatory solution of
Eq.~(\ref{9}), $x_1(t)\sim\text{exp} ({i}\Omega t)$ becomes
unstable, that is when $\text{Im}\,\Omega<0$. The smallness of the
r.h.s. of Eq.~(\ref{9}) allows one to obtain the imaginary part of
$\Omega$ by perturbation theory. The values of the parameters
$\Gamma_1, \Gamma_2$ and $h$ for which $\text{Im}\,\Omega<0$,
corresponding to the
shuttle regime of electron transport, satisfy the inequality
\begin{equation}\label{10}
h^4+A_1h^2+A_2<0,
\end{equation}
where
\begin{eqnarray}
A_1&=&-3\left(\Gamma_1^2+1\right)-\frac{\Gamma_1^2
-\Gamma_2^2}{8\Gamma_1^2}\left(5\Gamma_1^2+4\right),\\
A_2&=&\frac{\left(\Gamma_1^2-\Gamma_2^2\right)
\left(\Gamma_1^2+4\right)}{8\Gamma_1^2}\left(\Gamma_1^2+1+
\frac{\Gamma_1^2-\Gamma_2^2}{4}\right).
\end{eqnarray}
%
Note that in the absence of an external magnetic field the
nanoelectromechanical coupling results in additional damping of the
mechanical subsystem (since the coefficient $A_2>0$).

It is evident from Eq.~(\ref{10}), which is biquadratic in the magnetic field ($h$),
that a shuttle instability occurs in a finite interval
of magnetic fields, $h_{\text{min}}<h<h_{\text{max}}$. Now we
introduce the degree of spin polarization as
\begin{equation}\label{100}
\eta=\frac{\vert N_\uparrow-N_\downarrow\vert}{N_\uparrow+
N_\downarrow}\simeq\frac{\Gamma-\gamma}{\Gamma+\gamma}
\end{equation}
where $N_\sigma$ is the number of particles with spin projection
$\sigma$. The last relation in Eq.~(\ref{100}) comes from the
definition of tunneling rates and our assumption that tunneling
amplitudes do not depend on spin projections. We analyze
Eq.~(\ref{10}) in two limiting cases: (i) $\Gamma\gg 1$, $1-\eta\ll 1$,
and (ii) $\Gamma\ll 1$, $1-\eta\ll 1$.

Note that in the limit of weak polarization, $\eta\rightarrow 0$,
magnetic forces are small, $J\rightarrow 0$, and electron shuttling
is supported by Coulomb forces. This case was considered in
Ref.~\onlinecite{ilin}, where  it was shown that in the limit of
weak polarization a magnetic field ceases to influence single electron
shuttling.

\begin{figure}
\centering
\includegraphics[width=0.85\columnwidth]{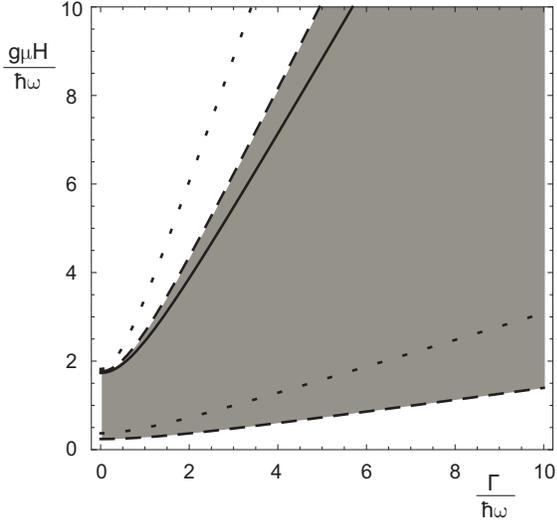}
\caption {Upper critical magnetic field (upper set of curves) and
lower critical field (lower set of curves) for the shuttle
transport regime plotted as a function of the normalized tunneling
rate $\Gamma/\hbar\omega$ of majority spin electrons for different
values of spin polarization $\eta$. The solid curves were plotted
for $\eta =1$ [100\% spin polarization, here the lower curve
coincides with the $x$-axis], the dashed curves were plotted for
$\eta = 0.8$, and the short-dashed curves for $\eta = 0.3$. The
shuttle regime corresponds to the area between the lower and upper
critical fields (dark region for the case of $\eta = 0.8$), while
outside this area one is in the vibronic regime.}
\end{figure}

In the adiabatic limit, (i), one finds (reverting to using
parameters with dimensions) that the
shuttle instability region is defined by the double inequality
\begin{equation}\label{11}
\frac{\Gamma}{2}\sqrt{\frac{1-\eta}{3}}<g\mu H<\sqrt{3}\,\Gamma.
\end{equation}
In the antiadiabatic limit, (ii), the critical magnetic fields are
determined by the vibron energy $\hbar\omega$ rather than the
tunneling rate $\Gamma$, so that
\begin{equation}\label{12}
\hbar\omega\sqrt{\frac{1-\eta}{3}}<g\mu H<\sqrt{3}\,\hbar\omega.
\end{equation}

For the general case that part of the $\Gamma, H$ parameter space
which corresponds to a magnetic shuttle instability is shown in
Fig.~1 for different degrees of spin polarization.
For a given (normalized) tunneling coupling $\Gamma/\hbar\omega$ 
the range of magnetic fields for which an instability occurs is shifted 
towards higher fields as the spin polarization decreases.
The upper critical
field for high spin polarizations, $\eta\rightarrow 1$, depends
linearly on $\eta$,
\begin{equation}\label{13}
g\mu
H_{\text{max}}\sim\sqrt{3\left(\Gamma^2+(\hbar\omega)^2\right)}
\left[1+K(1-\eta)\right]\,,
\end{equation}
where $K\sim 1$ is a positive constant.

The dependence of the lower critical magnetic field on $\eta$ is
weaker. In the antiadiabatic regime one finds that
\begin{equation}\label{14}
g\mu H_{\text{min}}\sim
\hbar\omega\sqrt{1-\eta},\quad \Gamma\ll\hbar\omega, 
\end{equation}
and
hence $H_{\text{min}}$ rapidly saturates to a constant value of
order $\hbar\omega/g\mu$ with decreasing spin polarization (see
Fig.~2, short-dashed curve).
In the adiabatic regime, 
$\Gamma\gg\hbar\omega$, the
lower critical field 
decreases linearly with increasing spin polarization, except in
the close vicinity of complete spin polarization, where
$H_{\text{min}} \propto \sqrt{1-\eta}$ (compare the solid curve in
Fig.~2).

\begin{figure}
\centering
\includegraphics[width=0.85\columnwidth]{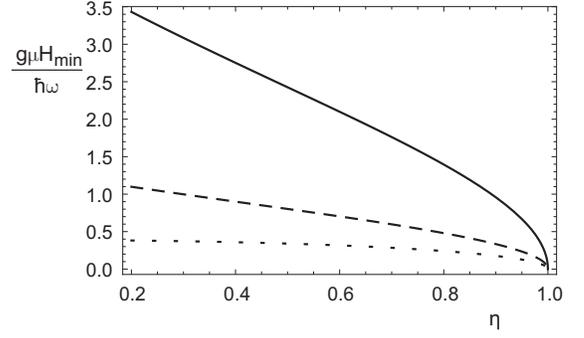}
\caption {Plots of the lower critical magnetic field
$H_{\text{min}}$, which defines the border between the vibronic
regime ($H<H_{\text{min}}$) and the shuttle transport regime
($H>H_{\text{min}}$; compare Fig.~1) vs. spin polarization $\eta$
for different values of the normalized tunneling rate
$\Gamma/\hbar\omega$ of majority spin electrons (solid curve:
$\Gamma/\hbar\omega=10$;  dashed curve: $\Gamma/\hbar\omega=3$;
short-dashed curve: $\Gamma/\hbar\omega=0.1$).}
\end{figure}

The appearance of an upper and a lower critical magnetic field has a
simple physical explanation. When $\mu H$ is the largest energy
scale in our problem, $\mu H\gg \Gamma,\,\hbar\omega$, the fast
precession of the electron spin of the dot in a perpendicular external magnetic field
nullifies the average spin and the magnetic shuttle instability
disappears. To estimate the upper field one may compare the
characteristic spin precession frequency, $\mu
H_{\text{max}}/\hbar$, with the electron tunneling rate,
$\Gamma/\hbar$, or the frequency of vibrations $\omega$. That is
$\mu H_{\text{max}}\sim\text{max}\left(\Gamma, \hbar\omega\right)$.
The lower critical field can be readily estimated for 
a high degree of spin polarization, $1-\eta\ll 1$. 
In this case we have to compare the average time between spin
flips, $\tau_f$, induced by a constant magnetic field $H$ in the
presence of an electron tunneling coupling $\Gamma$ with the
characteristic life-time of minority spin electrons on the dot,
$\sim \hbar/\gamma$. The spin-flip rate $\nu_f$ in weak magnetic
fields $H$ can be estimated by perturbation theory
with the result that $\hbar\nu_f\sim(\mu H)^2/\text{max}(\Gamma,
\hbar\omega)$. Therefore the lower magnetic field is strongly
sensitive to spin polarization,
\begin{equation}\label{15}
\mu H_{\text{min}}\sim
\sqrt{\Gamma\gamma}\,\text{max}(\Gamma,\hbar\omega)\sim
\sqrt{1-\eta}\,\text{max}(\Gamma,\hbar\omega),
\end{equation}
and disappears for 100\% spin-polarized electrons ($\eta=1)$.

Next we estimate the maximum rate of (exponential) increase,
$r_m=-\text{Im}\{\Omega (H_{\text{opt}})\}$, of the QD oscillation
amplitude in the shuttle regime.
In the adiabatic limit, $\Gamma\gg \hbar\omega$,
one finds that $g\mu H_{\text{opt}}\simeq 0.4\, \Gamma$ and that
\begin{equation}\label{16}
r_m\simeq C\frac{\omega J}{\Gamma}\left(\frac{x_0}{l}\right)^2,
\end{equation}
where $C\sim 0.1$ is a small numerical factor. In the case
$\Gamma\ll \hbar\omega$, which we are interested in here, the
maximum rate is realized when $g\mu H_{\text{opt}}\simeq
\hbar\omega$, corresponding to
\begin{equation}\label{17}
r_m\simeq
\frac{\Gamma}{\hbar}\frac{J}{\hbar\omega}\left(\frac{x_0}{l}\right)^2\,,
\end{equation}
where we omit a numerical factor of the order of one.

In the presence of dissipation in the mechanical subsystem, which
can be described by adding a phenomenological friction term
$\gamma_d \dot{x}_c(t)$ to the equation of motion (\ref{8})
($\gamma_d=\omega/Q$, where $Q$ is the quality factor), the
shuttling regime appears when $r_m>\omega/Q$. Therefore electron
shuttling in a $C_{60}$-based molecular transistor with magnetic
electrodes could be realized if the quality factor $Q$ of the
mechanical resonator obeys the inequality
\begin{equation}
\label{18}
Q>Q_{\text{opt}}=\frac{\left(\hbar\omega\right)^2}{J\Gamma}
\left(\frac{l}{x_0}\right)^2.
\end{equation}

For the experimental setup in Ref.~\onlinecite{park}, where
fullerene vibrations were observed, the factor
$\left(l/x_0\right)^2\simeq 10^3$ and
$\Gamma\ll\hbar\omega\sim$~5~meV (one can estimate
$\Gamma\sim$~0.1 -- 0.5~meV from the maximal current measured in
Ref.~\onlinecite{park}). In the $C_{60}$-based transistor with
magnetic (Ni) leads $J\sim\Gamma\sim$~10~meV (see
Ref.~\onlinecite{pasupathy}). From Eq.~(\ref{18}) one can estimate
that the required quality factor is $Q\geq 10^3-10^4$. However the
optimal external magnetic field in this case,
$H_{\text{opt}}\simeq$~50~T, is too high. Instead, we therefore
estimate $Q$ for magnetic fields in the vicinity of the lower
critical magnetic field $H\geq H_{\text{min}}$ where magnetic
fields for a very high degree of electron spin polarization ($\sim
99\%$) could be of the order of a few tesla. In this case
($\hbar\omega\gg \Gamma$, $1-\eta\ll 1$) 
\begin{equation}\label{19}
r(\eta)\simeq \omega\frac{J \Gamma
(1-\eta)}{\Gamma^2+4(1-\eta)(\hbar\omega)^2}\left(\frac{l}{x_0}\right)^2.
\end{equation}
Assuming that $\Gamma\simeq\sqrt{1-\eta}\,\hbar\omega$ we find
$Q\sim Q_{\text{opt}}/(1-\eta)$.

\section{4. Conclusions}
In summary we have considered the feasibility of observing magnetically
driven single-electron shuttling under 
realistic conditions 
corresponding to an already experimentally realized $C_{60}$-based
single-molecule transistor with magnetic leads. The main
requirement for magnetic shuttling is the presence of an external
magnetic field that induces electron
spin flips. We have shown that the optimal magnetic field, defined
as the field that maximizes the rate of increase of the shuttling
amplitude, is determined by the vibration frequency $\omega$. For
fullerene-based single-electron transistors this frequency could
be in the THz region \cite{park} 
with 
corresponding
magnetic fields 
in the region of several tenths of teslas. For magnetic electrodes
with a very high degree of spin polarization one needs less strong
(by an order of magnitude) magnetic fields. However, the quality
factor of the corresponding mechanical resonator has to be
exceptionally high, $Q\geq 10^5$.

{\bf Acknowledgements:} Financial support from the Leading Foreign
Research Institutes Recruitment Program (2009-00514) of NRF,
Korea, and the Swedish Research Council (VR) is gratefully
acknowledged. OI, IK and SK acknowledge financial support from
National Academy of Science of Ukraine, Grant No 4/15--N. IK and
SK thank the Department of Physics at the University of Gothenburg
and the Department of Physics and Astronomy at Seoul National
University for their hospitality.


\begin{thebibliography}{99}
\bibitem{pulkin}
R.~I.~Shekhter, A.~Pulkin, and M.~Jonson, Phys. Rev. B {\bf 86}, 100404(R)
(2012).

\bibitem{galperin}
M.~Galperin, M.~A.~Ratner, and A.~Nitzan, J. Phys.: Condens. Matter {\bf
19}, 103201 (2007).

\bibitem{krive}
I.~V.~Krive, A.~Palevski, R.~I.~Shekhter, and M.~Jonson, Low
Temp. Phys. {\bf 36}, 119 (2010).

\bibitem{park}
H.~Park, J.~Park, A.~K.~L.~Lim, E.~H.~Anderson,  A.~P.~Alivisatos, and
P.~L.~McEuen, Nature {\bf 407}, 57 (2000).

\bibitem{shekhter}
D.~Fedorets, L.~Y.~Gorelik, R.~I.~Shekhter, and M.~Jonson, Europhys. Lett.
{\bf 58}, 99 (2002).

\bibitem{flensberg}
S.~Braig and K.~Flensberg, Phys. Rev. B {\bf 68}, 205324 (2003).

\bibitem{pasupathy}
A. N. Pasupathy, R.~C.~Bialczak, J.~Martinek, J.~E.~Grose, L.~A.~K.~Donev,
P.~L.~McEuen, and D.~C.~Ralph, Science {\bf 306}, 86 (2004).

\bibitem{kul}
S.~I.~Kulinich, L.~Y.~Gorelik, A.~N.~Kalinenko, I.~V.~Krive, R.~I.~Shekhter,
Y.~W.~Park, and M.~Jonson, Phys. Rev. Lett. {\bf 112}, 117206 (2014).

\bibitem{jonson}
D.~Fedorets, L.~Y.~Gorelik, R.~I.~Shekhter, and M.~Jonson, Phys. Rev. Lett.
{\bf 95}, 057203 (2005).

\bibitem{gorelik} L. Y. Gorelik, D.~Fedorets, R.~I.~Shekhter,
and M.~Jonson, New J. Phys. {\bf 7}, 242 (2005).

\bibitem{parafilo}
S.~I.~Kulinich, L.~Y.~Gorelik, A.~V.~Parafilo, R.~I.~Shekhter,Y.~W.~Park, and
M.~Jonson, Low Temp. Phys. {\bf 40}, 907 (2014).

\bibitem{ilin}
O.~A.~Ilinskaya, S.~I.~Kulinich, I.~V.~Krive, R.~I.~Shekhter, and M.~Jonson,
Low Temp. Phys. {\bf41}, 70 (2015).

\bibitem{mis}
Notice that in Eq.~(8) of Ref.~\onlinecite{ilin} (analogous to
Eq.~(\ref{baseq2}) of the present paper) there are misprints. The
corrections are as follows - minus sign in the $h$-term and the
terms $\Gamma_S^\uparrow(x)$ and $\Gamma_S^\downarrow(x)$ have to
be omitted.
\end{thebibliography}
\end{document}